\documentclass[10pt,conference,compsocconf,letter]{IEEEtran}

\usepackage{cite}
\usepackage[pdftex]{graphicx}
\usepackage{amsfonts,amssymb,amsmath,alltt}
\usepackage{array}
\usepackage{mdwmath}
\usepackage{mdwtab}
\usepackage{fixltx2e}
\usepackage{stfloats}
\usepackage{url}
\usepackage{xspace}
\usepackage{color}
\usepackage{stmaryrd}

\usepackage{alltt}

\newenvironment{ttbox}{\begin{alltt}\ttbraces\small\tt}%
                      {\end{alltt}}
\def\ttbraces{\let\.=\nobreak\chardef\{=`\{\chardef\}=`\}\chardef\|=`\\}

\newcommand\ttexists{\mbox{{$\exists$}}}

\newcommand\ttin{\mbox{{$\in$}}}

\newcommand\ttrelI{\mbox{{$\to_{i}$}}}
\newcommand\ttrelIstar{\mbox{{$\to_{i}^*$}}}

\newcommand\ttvdash{\mbox{{$\vdash$}}}

\newcommand\ttsigma{\mbox{{$\sigma$}}}

\newcommand{\ttcalN}[1]{\mbox{{${\cal{N}}_{\texttt{#1}}$}}} 
\newcommand\ttattand[1]{\mbox{{$\oplus_{\wedge}^{#1}$}}}
\newcommand\ttattor[1]{\mbox{{$\oplus_{\vee}^{#1}$}}}

\begin{document}
\title{Explanation by Automated Reasoning Using the Isabelle Infrastructure Framework}
\author{\IEEEauthorblockN{Florian Kamm\"uller}
\IEEEauthorblockA{Department of Computer Science\\Middlesex University London\\
f.kammueller@mdx.ac.uk}
}
\maketitle              

\begin{abstract}
  In this paper, we propose the use of interactive theorem proving for explainable machine learning.
  After presenting our proposition, we illustrate it on the dedicated application of explaining security
  attacks using the Isabelle Infrastructure framework and its process of dependability engineering. This
  formal framework and process provides the logics for specification and modeling. Attacks on security of
  the system are explained by specification and proofs in the Isabelle   Infrastructure framework.
  Existing case studies of dependability engineering in Isabelle are used as feasibility studies to
  illustrate how different aspects of explanations are covered by the Isabelle Infrastructure framework.
\end{abstract}  

\section{Proposing Interactive Theorem Proving for Explainable Machine Learning}
\label{sec:intro}
Machine Learning (ML) is everywhere in Computer Science now. One may almost say that all of
Computer Science has now become a part of ML and is viewed as a technique within the greater
realm of Data Science or Data Engineering. But while this major trend like many other trends
prevails, we should not forget that Artificial Intelligence (AI) is the original goal of what was the
starting point of machine learning and that Automated Reasoning has been created as a means to
provide for artificial intelligent systems a mechanical way of imitating human reasoning by implementing
logics and automatizing proof.

When we think of how to explain why a specific solution for a problem is a solution, the purest way
to do so is to explain it by way of mathematically precise arguments -- which is equivalent to
providing a logically sound proof in a mathematical model of the solution domain or context. An ML
algorithm would do the same, for example, by providing a decision tree to explain a solution, but usually
the ML explanations which are generated by the ML model itself are very close to the ML implementation.
So, they often fail to give a satisfactory, i.e. human understandable explanation.

This paper shows our point of view on a tangible way forward to combining interactive theorem
proving with machine learning (ML). Different from the main stream of using ML to improve automated
verification, we propose an integration at a higher level, using logical modeling and automated
reasoning for explainability of machine learning solutions.
The main idea of our proposal is based on one major fact about logic and proof:
\begin{quote}{\it 
Reasoning is not only a very natural
way of explanation but it is also the most complete possible one since it provides a mathematical
proof on a formal model.}
\end{quote}
In the spirit of this thought, we provide a proof of concept on a framework that has been
established for security and privacy analysis, the Isabelle Infrastructure framework.
In this paper, we thus first introduce this framework by summarizing its basic conepts and
various applications (Section\ref{sec:infra}).
After contrasting to some other conceptual approaches to ML and theorem proving
including explanation (Section \ref{sec:rel}),
we briefly sketch our conceptual proposal (Section \ref{sec:prop})

\section{Isabelle Infrastructure Framework}
\label{sec:infra}
The Isabelle Infrastructure framework is implemented as an instance of
Higher Order Logic in the interactive generic theorem prover Isabelle/HOL \cite{npw:02}.
The framework enables formalizing and proving of systems with physical and logical components,
actors and policies. It has been designed for the analysis of insider 
threats. However, the implemented theory of temporal logic combined with Kripke structures and its
generic notion of state transitions are a perfect match to be combined with  attack trees into
a process for formal security engineering \cite{suc:16} including an accompanying framework \cite{kam:19a}.

\subsubsection{Kripke structures, CTL and Attack Trees}
A number of case studies 
have contributed to shape the Isabelle framework into a general framework for
the state-based security analysis of infrastructures with policies and actors. Temporal logic
and Kripke structures are deeply embedded into Isabelle's Higher Order logic thereby
enabling meta-theoretical proofs about the foundations: for example, equivalence between attack trees 
and CTL statements have been established \cite{kam:18b} providing sound foundations for applications.
This foundation provides a generic notion of state transition on which attack trees and
temporal logic can be used to express properties for applications. 
The logical concepts and related notions thus provided for sound application modeling are:
\begin{itemize}
\item {Kripke structures and state transitions:}\\ 
A generic state transition relation is $\ttrelI$; Kripke structures
over a set of states \texttt{t} reachable by $\ttrelI$ from an initial 
state set \texttt{I} can be constructed by the \texttt{Kripke} constructor as
\begin{ttbox}
Kripke \{t. \ttexists i \ttin I. i \ttrelIstar t\} I
\end{ttbox}
\item {CTL statements:}\\ 
We can use the Computation Tree Logic (CTL) to specify dependability properties as 
\begin{ttbox}
K \ttvdash {\sf EF} s
\end{ttbox}
This formula states that in Kripke structure \texttt{K} there is a path ({\sf E}) on which
the property \texttt{s} (given as the set of states in which the property is true) will eventually ({\sf F}) hold.
\item {Attack trees:} \\
attack trees are defined as a recursive datatype in Isabelle having three constructors: 
$\oplus_\vee$ creates or-trees and $\oplus_\wedge$ creates 
and-trees.
And-attack trees $l \ttattand s$ and or-attack trees $l \ttattor s$ 
consist of a list of sub-attacks which are themselves recursively given as attack trees.
The third constructor takes as input a pair of state sets constructing a base attack step between
two state sets. For example, for the sets \texttt{I} and \texttt{s} this is written as \texttt{\ttcalN{(I,s)}}.
As a further example, a two step and-attack leading from state set \texttt{I} via \texttt{si} to \texttt{s} is
expressed as 
\begin{ttbox}
\ttvdash [\ttcalN{(I,si)},\ttcalN{(si,s)}]\ttattand{\texttt{(I,s)}}
\end{ttbox}
\item {Attack tree refinement, validity and adequacy:}\\
Attack trees can be constructed also by a refinement process but this differs from
the system refinement presented in the paper \cite{kam:21a}.
An abstract attack tree may be refined by spelling out the attack steps until a valid attack
is reached:

\texttt{\ttvdash A :: (\ttsigma :: state) attree)}.

The validity is defined constructively so that code can be generated from it.
Adequacy with respect to a formal semantics in CTL is proved and can be used to
facilitate actual application verification. This is used for the stepwise system refinements
central to the methodology called Refinement-Risk cycle developed for the Isabelle Infrastructure
framework \cite{kam:21a}.
\end{itemize}

A whole range of publications have documented the development of the Isabelle Insider framework.
The publications \cite{kp:13,kp:14,kp:16} first define the fundamental notions of insiderness, policies,
and behaviour showing how these concepts are able to express the classical insider threat patterns
identified in the seminal CERT guide on insider threats \cite{cmt:12}.
This Isabelle Insider framework has been applied to auction protocols \cite{kkp:16,kkp:16a} illustrating
that the Insider framework can embed the inductive approach to protocol verification \cite{pau:98}.
An Airplane
case study \cite{kk:16,kk:21} revealed the need for dynamic state verification leading to 
the extension of adding a mutable state. Meanwhile, the embedding of Kripke structures and CTL
into Isabelle have enabled the emulation of Modelchecking and to provide a semantics for attack
trees \cite{kam:17a,kam:17b,kam:17c,kam:18b,kam:19a}.
Attack trees have provided the leverage to integrate Isabelle formal reasoning for IoT systems
as has been illustrated in the CHIST-ERA project SUCCESS \cite{suc:16} where 
attack trees have been used in combination with  the Behaviour Interaction Priority (BIP) component 
architecture model to develop security and privacy enhanced IoT solutions.
This development has emphasized the technical rather than the psychological side of the framework
development and thus branched off the development of the Isabelle {\it Insider} framework into the
Isabelle {\it Infrastructure} framework. Since the strong expressiveness of Isabelle allows to formalize
the IoT scenarios as well as actors and policies, the latter framework can also be applied to 
evaluate IoT scenarios with respect to policies like the European data privacy regulation
GDPR \cite{kam:18a}. Application to security protocols first pioneered in the
auction protocol application \cite{kkp:16,kkp:16a} has further motivated the analysis of Quantum Cryptography
which in turn necessitated the extension by probabilities \cite{kam:19b,kam:19c,kam:19d}.

Requirements raised by these various security and privacy case studies have shown the need for a
cyclic engineering process for developing specifications and refining them towards implementations.
A first case study takes the IoT healthcare application and exemplifies a step-by-step
refinement interspersed with attack analysis using attack trees to increase privacy by ultimately
introducing a blockchain for access control \cite{kam:19a}.
First ideas to support a dedicated security refinement process are available in a preliminary
arxive paper \cite{kam:20a} but only the follow-up publication \cite{ka:21} provides the first
full formalization of the RR-cycle and illustrates its application completely on the Corona-virus
Warn App (CWA). The earlier workshop publication \cite{kl:20} provided the formalisation of the
CWA illustrating the first two steps but it did not 
introduce the fully formalised RR-cycle nor did it apply it to arrive at a solution satisfying the
global privacy policy \cite{kam:21a}. 

\section{Machine Learning, Explanation and Theorem Proving}
\label{sec:rel}
If theorem proving could automatically be solved by machine learning, we would solve the P=NP problem
\cite{wk:19}. Nevertheless, ML has been successfully employed within theorem provers to enhance the
decision processes. Also in Isabelle, the sledgehammer tool uses ML mainly to select lemmas.

A very relevant work by Vigano and Magazzeni \cite{vm:20} focuses the idea of explainability
on security, coining the notion of {\it XSec} or {\it Explainable Security}.
The authors propose a full new research programme for the notion of explainability in security
in which they identify the ``Six Ws'' of XSec: Who? What? Where? When? Why? And hoW?
They position their paper clearly into the context of some earlier works along the same lines,
e.g. \cite{bkg:14,pie:11}, but go beyond the earlier works by extending the scope and presenting
a very concise yet complete description of the challenges.
As opposed to XAI in general, the paper shows how already in understanding explanations only for
the focus area of security (as opposed to all application domains of IT) is quite a task. Also they
point out that XAI is merely concerned with explaining the technical solution provided by ML, whereas
XSec looks at various other levels most prominently, the human user, by addressing domains like
{\it usable security} and {\it security awareness}, and {\it security economics} \cite{vm:20}[p. 294].

Our point of view is quite similar to Vigano's and Magazzeni's but we emphasize the technical side of
explanation using interactive theorem proving and the Isabelle Infrastructure framework, while they
focus on differentiating the notion of explanation from different aspects, for example, stake holders,
system view, and abstraction levels.
However, the notion of refinement defined for the process of dependability engineering for the Isabelle
Infrastructure framework \cite{kam:21a} allows addressing most of the Six Ws, because our model
includes actors and policies and allows differentiation between insider and outsider attacks, expression
of awareness \cite{ka:21}.
Thus, we could strictly follow the Ws when explaining our proposition but we believe it is better to
contemplate the Ws simply in the context of classical Software Engineering that has similar Ws.
Moreover, the Refinement-Risk cycle of dependability engineering can be seen as specification refinement
framework that employs the classical AI technique of automated reasoning. Surely, the human aspect
versus the system aspect on the Sic Ws of XSec brings in various different view points but these are
inherent in if the contexts, that are needed
for the interpretation are present in the model. Otherwise, they simply have to be added to it,
for example, by using refinement to integrate these aspects of reality into the model. Then the
Isabelle Infrastructure framework allows explanation for various purposes, audiences, technical
levels (HW/SW). policies, localities  and other physical aspects. Thus, we can answer all Six Ws
and argue that is what human centric software, security, and dependability engineering are all about.

Moreover, despite contrasting from the approach by Vigano and Megazzini, we follow the classical
engineering approach of Fault-tree analysis, more concretely using Attack Trees, and propose a dual
process of attack versus security protection goal analysis which in itself offers a direct input to
ML, for example to produce features that could be used for Decision trees as well as metrics that
could provide feedback for optimization techniques as used in reinforcement learning.

\section{Explaining (not only) Security by the Isabelle Infrastructure framework}
\label{sec:prop}
This section describes the core ideas of explanation provided by applying the Isabelle
Infrastructure framework.
\subsection{State transition systems and attack trees as a dual way of explanation}
One important aspect of explanation that is not restricted to security at all is to provide a
step by step trace of state transitions to explain how a specific state may appear. This can explain
where a problem lies, for example, to explain how an ML algorithm arrived at a decision for a medical
diagnosis by lining up a number of steps that lead to it.

In the Isabelle Infrastructure framework the notion of state transition systems is provided as a generic
theory based on Kripke structures to represent state graphs over arbitrary types of states
and using the branching time logic CTL to express temporal logical formulas over them.
The correspondance between the CTL formulas of reachability and attack trees and the proof of
adequacy are suitable to allow for a dual step by step analysis of a system dove-tailing the fault
analysis with a specification refinement. This dove-tailing process leads to an elaborate process not
only of explaining faults of system designs and how they can be reached practically by a series of
actions but also an explanation of additional features of a system that are motivated by the detected
fault. For example, when it comes to human awareness and usable security an explanation of a necessary
security measure that is imposed on a user can be readily illustrated by an attack graph or its
equivalent attack path that can be readilty produced by the adequacy theory.

\subsection{Human and Locality Aspects}
The Isabelle Infrastructure framework has initially been designed to be merely focused on
modeling and analyzing Insider threats before it became extended into what is now known as
the Isabelle Infrastructure framework. Due to this initial motivation the framework explicitly supports
the notion of human actors within networks of physical and virtual locations. These aspects are important
to model various different stake holders to enable explanations to different audiences having different
view points and needing different levels of detail and complexity in their explanations. For example,
the explanation of a security threat will have a substantially different form if produced for a security
analyst of to a system end user.
Due to the explicit representation of human actors as well as their locations and other variable features,
the Isabelle Infrastructure framework supports a fine grained control over the definition of 
applications thus enabling very flexible support of explanation about human aspects and suited to human
understanding.

Also the human aspect necessitates consideration of the human condition, in particular psychological
characterizations. The Isabelle Infrastructure framework, by augmenting the Isabelle Insider framework,
provides for such characterization. For example, when considering insiderness, the state of the insider
is characterized by a predicate that allows to use this state within a logical analysis of security
and privacy threats to a system. Although these characterizations are axiomatic in the sense that the
definition of the insider predicate is based on empirical results that have been externally input into
the specification, it is in principle feasible to enrich the cognitive model of the human in the
Isabelle Insider framework. A first step towards that has been done by experimenting with an extension
of a notion of human awareness to support additionally analysis of unintentional insiders for 
human unawareness of privacy risks in social media \cite{ka:21}.

\subsection{Dependability  Engineering: Specifying Protection Goals and Quantifying Attackers}
The process of Dependability Engineering -- the Refinement-Risk (RR) cycle -- conceived for the
Isabelle Infrastructure framework \cite{kam:20a} allows a human centric system specification to be
refined step-by-step following an iteration of finding faults within a system specification and
refining this specification by more sophisticated data types or additional rules or changes to the
semantics of system functions. The data type refinement allows integrating  for example, more
restrictive measures to control data, for example, using blockchains to enhance data consistency,
or data labeling for access controzl. This refinement is triggered by previously found flaws in the
system and thus provides concrete motivations for such design decisions leading to constructive
explanations. Similarly, additional constraints on rules that are introduced in a refinement step
of the RR-cycle are motivated by previously found attacks, for example, the necessity to change
the ephemeral id of every user when they move to a new location instantaneously at moving time for the
Corona-Warn-App is motivated by an identification attack \cite{kl:20,kam:21a}.

Since the RR-cycle is based on the idea of refinement, another requirement for a flexible explanation
comes in for free: if we want to explain to different audiences or at different technical levels, we
equally need to refine (or abstract) definitions of data-types, rules for policies, or descriptions
of algorithms. The Isabelle Infrastructure framework directly supports these expressions at different
abstraction levels and from different view points.

\subsection{Quantification}
An important aspect is quantification for explanation. Very often an explanation will not be possible
in a possibilistic way. A quantification could be given by adding probabililies as well as other
quanitative data, like costs, to explanations. For example, for a security attack the cost that
an attacker is estimated to invest maximally on a specific attack step is an inevitable ingredient
for a realistic attacker model. Simlarly, the likelihood of a successful attack of a certain attack
step could be needed for an analysis. Attack trees support these types of quantification. Naturally,
the Isabelle Infrastructure framework also supports them. The application to the security analysis
of Quantum Cryptography, i.e., the modeling and analysis of the Quantum Key Distribution protocol (QKD)
lead to the extension for probabilistic state transition systems \cite{kam:19b,kam:19c,kam:19d}.

Quantification can also be a useful explanation for the process of learning for example
by quantifying a distance to an attack goal. In that sense, quantified explanation can be a useful
feedback for machine learning itself.


\section{Conclusions}
In this paper we have proposed the sue of Automated Reasoning in the particular instance of the
Isabelle Infrastructure framework for Explanation. 
We summarized the work that lead to the creation of the Isabelle Infrastructure framework highlighting
the existing applications and extensions. After studying some related work on explanation, we provided
a range of conceptual points that argued why and how the Isabelle Infrastructure framework supports
explanation.

\bibliographystyle{IEEEtran} \bibliography{../insider}

\end{document}